\documentclass[11pt]{article}
\usepackage{epsfig}
\usepackage{amssymb}
\usepackage{latexsym}
\usepackage{helvet}

    \ifx\proof\undefined
    \newenvironment{proof}{
    \smallskip
    \noindent\emph{Proof.}}{\hfill\(\Box\)
    \bigskip
    } \fi

\newcommand{\Real}{\mathbb R}
\newcommand{\Proj}{\mathbb P}

\newcommand{\FigWidth}{4.8}

\def\ni{\noindent}

\newcommand {\secSpc} {\vskip 0cm}
\newcommand {\subSecSpc} {\vskip 0cm}
\newcommand {\topFigSpc} {\vskip 0cm}
\newcommand {\capFigSpc} {\vskip 0cm}
\newcommand {\butFigSpc} {\vskip 0cm}
\title{  New Visualization of Surfaces in
Parallel Coordinates \\ -- Eliminating Ambiguity and Some
``Over-Plotting''}
\author{
{Zur Izhakian}  \\
{Department of Computer Science} \\
{Tel Aviv University} \\
{Ramat Aviv, 69978} \\
{Tel Aviv, Israel}\\
{email: zzur@post.tau.ac.il}\\
}

\date{February 2004}

\begin{document}
\maketitle

%******************************* abstract *********************************
\begin{abstract}
$\cal{A}$ point $P \in \Real^n$ is represented in Parallel
Coordinates by a polygonal line $\bar{P}$ (see \cite{Insel99a} for
a recent survey). Earlier \cite{inselberg85plane}, a surface
$\sigma$ was represented as the {\em envelope} of the polygonal
lines representing it's points. This is ambiguous in the sense
that {\em different} surfaces can provide the {\em same}
envelopes.  Here the ambiguity is eliminated by considering the
surface $\sigma$ as the envelope of it's {\em tangent planes} and
in turn, representing each of these planes by $n$-$1$ points
\cite{Insel99a}. This, with some future extension, can yield a new
and unambiguous representation, $\bar{\sigma}$, of the surface
consisting of $n$-$1$ planar regions whose properties correspond
lead to the {\em recognition} of the surfaces' properties i.e.
developable, ruled etc. \cite{hung92smooth}) and {\em
classification} criteria.

It is further shown that the image (i.e. representation) of an
algebraic surface of degree $2$ in $\Real^n$ is a region whose
boundary is also an algebraic curve of degree $2$. This includes
some {\em non-convex} surfaces which with the previous ambiguous
representation could not be treated. An efficient construction
algorithm for the representation of the quadratic surfaces (given
either by {\em explicit} or {\em implicit} equation) is provided.
The results obtained are suitable for applications, to be
presented in a future paper, and in particular for the
approximation of complex surfaces based on their {\em planar}
images. An additional benefit is the elimination of the
``over-plotting'' problem i.e. the ``bunching'' of polygonal lines
which often obscure part of the parallel-coordinate display.

\end{abstract}

%******************************* keywords *********************************
%\begin{keywords}
{\bf keywords}: Scientific Visualization and HMI.
%\end{keywords}

%******************************* AMS *********************************
%\begin{AMS}
{\bf AMS} : 76M27
%\end{AMS}

\section{INTRODUCTION}

\cal{O}ur purpose here is expository, sparing the reader from most
of the mathematical tribulations and, focusing on the more
intuitive aspects of the representational results. After a short
review of the fundamentals, the essentials of the mathematical
development are given together with some detailed examples to
clarify the nuances and satisfy the more mathematically inclined.

\begin{figure}[!h]
\centering
\includegraphics[width=3 in]{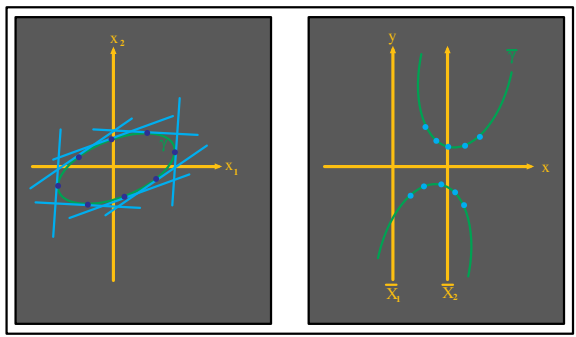}
\vskip -0.2cm \caption{\label{Curve} Point-curve mapped into
point-curve, result of considering the point-curve as the envelope
of it's tangents.}
\end{figure}

In parallel coordinates (abbr. $\|$-coords), a point in
$\Real^{2}$ is represented by a line and a line is represented by
a point yielding a fundamental $point \leftrightarrow line$
duality. There follows the representation of $p$-flats (planes of
dimension $2 \leq p \leq n-1$) in $\Real^{n}$ in terms of indexed
points \cite{inselberg85plane}. Naturally, for non-linear objects
the representation is more complex, especially if they are also
non-convex. The points of a curve in $\Real^{2}$ can be mapped
directly into a family of lines whose {\em envelope} defines a
curve (``line-curve''). Actually this is awkward and also clutters
the display. Instead we map the tangents of the original curve
into points to obtain the ``point-curve'', sometimes called
``dual-curve'', image directly as shown in Fig. \ref{Curve}. In
short, this approach provides a convenient  {\em point-to-point}
mapping \cite{Insel99a}.

%%%%%%%%%%%%%%%%%%%%%%%%%%%%%%%%%%%%%%%%%%%%%%%%%%%%%%%%%%%%%%%%%%%%
%%                      curve_pt2pt_eps
%\vskip -0.5cm
% \vskip -0.3cm
%%%%%%%%%%%%%%%%%%%%%%%%%%%%%%%%%%%%%%%%%%%%%%%%%%%%%%%%%%%%%%%%%%%%

Applying these considerations it was proved that the image of an
algebraic curve of degree $n$ is also algebraic of degree at most
$n(n - 1)$ in the absence of singular points \cite{zur:01:ms}.
This theorem is a generalization of the known result that conics
are mapped into conics \cite{dimsdale84conic} in six different
ways.

Perhaps we are ``pushing our luck'', our intent is to apply next
the point-to-point mapping in the representation of surfaces
considered as the envelope of their tangent planes; with the
resulting image being constructed from the representation of
tangent planes \cite{Insel99a}. As has already been pointed out,
planes can be represented in $\|$-coords by indexed points. The
collection of these planar points, grouped for each index, is the
representation of the surface.

%%%%%%%%%%%%%%%%%%%%%%%%%%%%%%%%%%%%%%%%%%%%%%%%%%%%%%%%%%%%%%%%%%%%
%%                   general_plane
\topFigSpc
\begin{figure*}%[!h]
\centering
\includegraphics[width=\FigWidth in]{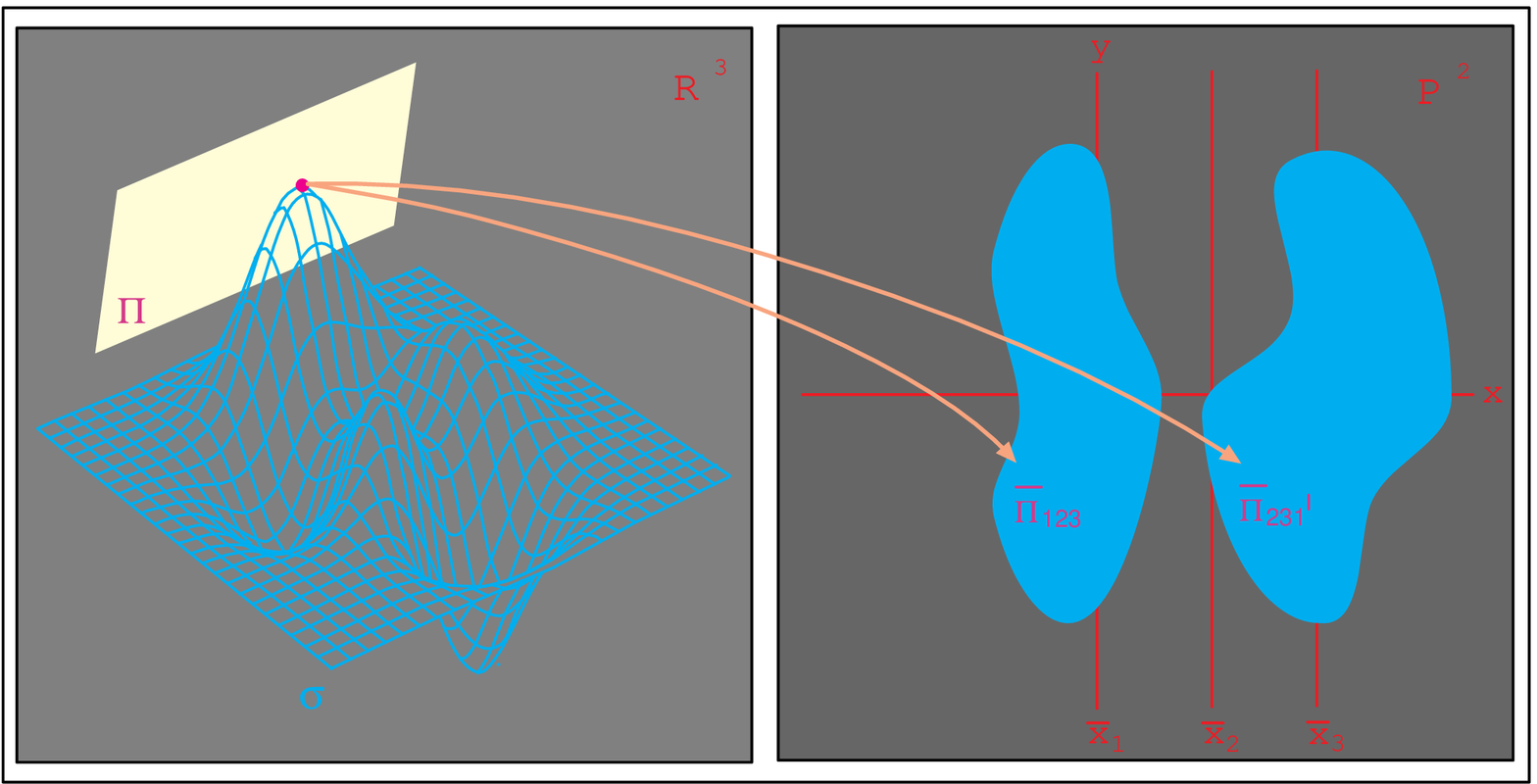}
\capFigSpc \caption{\label{figGenralPlane} For a hyper-surface,
it's representation's boundary curves are determined }\vskip
-0.3cm {from the images of curves contained in hyper-surface.}
\butFigSpc
\end{figure*}
%%%%%%%%%%%%%%%%%%%%%%%%%%%%%%%%%%%%%%%%%%%%%%%%%%%%%%%%%%%%%%%%%%%%

In the past \cite{inselberg85plane}, surfaces were represented in
$\|$-coords by the envelope of the polygonal lines representing
the surfaces' points. By itself this is ambiguous. For example the
image of a sphere in $n$-dimensions is the same as the image of
the surface obtained by the intersection of $n$ cylinders properly
aligned having the same radius. In applications this was
ameliorated by accessing the correct equation of the surface when
needed. Not only is the ambiguity completely removed with the new
representation, but also {\em non-convex} surfaces can be nicely
treated something that was not possible previously.

Hung \cite{hung92smooth} first applied this notion and found that
regions representing {\em developable} surfaces consists only of
the boundary curves (i.e. there are no interior points), and also
that {\em ruled} surfaces can be recognized from characteristic
properties of their corresponding regions. Encouraged by these
initial results the analysis is extended to more general surfaces
yielding useful criteria in the approximation of surfaces by
simpler ones; but we are getting ahead of ourselves.

At first we lay the foundations, then derive the representation of
quadratic algebraic surfaces and further generalize to higher
dimension as well as more complex hyper-surfaces. As a result an
efficient algorithm for constructing the representation of the
quadratic surfaces (given either by {\em explicit} or {\em
implicit} equation), and a proof that the image of an algebraic
surface of degree $2$ in $\Real^n$ is also an algebraic curve and
of degree $2$ are obtained.
%******************************* SECT 1 *********************************
\secSpc
\section{GENERAL REPRESENTATION OF HYPER-SURFACE}
%\section{General Representation of Hyper-Surfaces}
$\cal{I}$n general, the method employed below applies to the class
of smooth hyper-surfaces in $\Real^n$ having a unique tangent
hyper-plane at each point. Equivalently, each such hyper-surface
is the envelope of it's tangent hyper-planes. This is our point of
departure, for it enables us to represent each tangent hyper-plane
in $\|$-coords by $n-1$ indexed points \cite{Insel99a}. The
hyper-surface's representation consists of the $n-1$ points sets,
one for each index \cite{hung92smooth}. For the present we
restrict our attention to algebraic hyper-surfaces and in
particular those defined by quadratic polynomials. To simplify
matters, most of the analysis is done in 3-dimensional space but
in a way which points to the generalization for $\Real^{n}$.

%~~~~~~~~~~~~~~~~~~~~~~~~~~~~ SUB ~~ SECT 1 ~~~~~~~~~~~~~~~~~~~~~~~~~~~~~
\subSecSpc
\subsection{Hyper-Planes Representation}
An $n$-dimensional hyper-plane $\pi$ in $\Real^{n}$
\begin{equation}\label{planeSur}
  \pi :\sum_{i=1}^{n} c_{i}x_{i}=c_{0},
\end{equation}
is represented by the $n-1$ indexed points \cite{Insel99a}. For
our purposes only the first
\begin{equation}\label{planeRep}
\bar{\pi}_{1 \dots n} = (\sum_{i=1}^{n}(i-1)
c_{i},c_{0},\sum_{i=1}^{n}c_{i}).
\end{equation}
needs to be studied. The remaining $n-2$ points have similar form
differing only in the factor $(i-1)$ of the $c_i$. An important
property is that the horizontal distance between the $i$-adjacent
(in the indexing) points is the equal to the coefficient $c_i$;
from which the sequence of indexed points can be generated from
the coefficients or vice-versa.

%~~~~~~~~~~~~~~~~~~~~~~~~~~~~ SUB ~~ SECT 1 ~~~~~~~~~~~~~~~~~~~~~~~~~~~~~
\subSecSpc
\subsection{Hyper-Surfaces Representation}

Moving on to the representation of non-linear hyper-surfaces in
$\Real^{n}$ from their tangent hyper-planes.
% As explained earlier
% this yields the of the complete family of the hyper-surface's
% tangent hyper-planes which provide the representation of the
% hyper-surface.
Let $\sigma$ be a smooth $n$-dimensional hyper-surface generated
by the differentiable function $F(x_{1},\dots,x_{n})=0$, and an
arbitrary point $(x^{0}_{1}, \dots ,x^{0}_{n}) \in \sigma$. Then
the hyper-surface's tangent hyper-plane at this point is given by
:
$$\sum_{i=1}^{n}(x_{i}-x^{0}_{i})\frac{\partial{F}}{\partial{x_{i}}}(x^{0}_{1},\dots,x^{0}_{n})
=0 .$$
Taking the $x^{0}_{i}$ as parameters, the coefficients of the
tangent hyper-planes can written as a function of points which
satisfy the hyper-surface's equation. Namely, the family of
tangent hyper-planes of $\sigma$ is represented in homogenous
coordinates, by a collection of sets $\bar{\sigma}$ (see Fig.
\ref{figGenralPlane}), containing the indexed points representing
each member (i.e. hyper-plane) of the family. Each of the $n-1$
indexed set, $\bar{\sigma}_{j \dots n 1' \dots (j-1)'}$, consists
of the points with the same index.

 In the remainder the analysis is confined to the first indexed set
 $\bar{\sigma}_{1 \dots n}$ using a shorter notation $
 \bar{\sigma}$ defined as,
\begin{equation}\label{planeRep}
 \bar{\sigma}=
 \left\{ \left(P(\bar{x}), \; S(\bar{x}), \; Q(\bar{x}) \right) \; | \;
\bar{x} \in \sigma) \right\},
\end{equation}
where for $\bar{\sigma}_{1 \dots n}$  and the tuple $\bar{x} =
(x_{1},\dots ,x_{n})$ :

  \begin{eqnarray}{}\label{PQSgen}
    \nonumber P(\bar{x}) & = &
\sum_{i=1}^{n}(i-1)\frac{\partial{F}}{\partial{x_{i}}}  \\
    S(\bar{x}) & = & \sum_{i=1}^{n}x_{i}\frac{\partial{F}}{\partial{x_{i}}} \\
\nonumber Q(\bar{x})& = &
\sum_{i=1}^{n}\frac{\partial{F}}{\partial{x_{i}}}.
  \end{eqnarray}
In general, the representation in $\|$-coords is constructed via
the rational transformations
\begin{equation}\label{XYrationalTransform}
x= \frac{P(\bar{x})}{Q(\bar{x})} \; \; \; , \; \; \;
y=\frac{S(\bar{x})}{Q(\bar{x})}.
\end{equation}
\noindent The representation of all these hyper-planes transform
an $n$-dimensional hyper-surface into subsets of $\Proj^2$;
regions which are distinguished from each other by their indices.
The algorithm which constructs and describes these regions is
presented next.

%%%%%%%%%%%%%%%%%%%%%%%%%%%%%%%%%%%%%%%%%%%%%%%%%%%%%%%%%%%%%%%%%%%%
%%                      general_curve
\topFigSpc
\begin{figure*}%[!h]
\centering
\includegraphics[width=\FigWidth in]{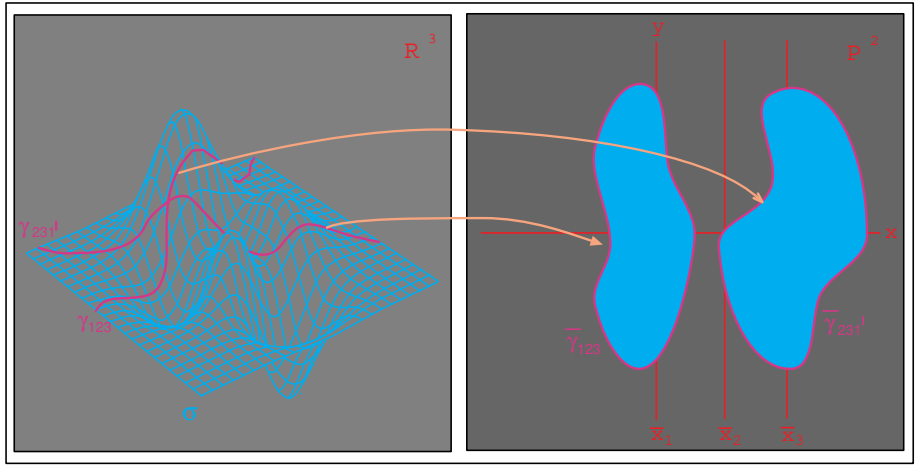}
\capFigSpc \caption{\label{figGenralCurve} The boundary of each
region is an image of a curve embedded in surface.} \butFigSpc
\end{figure*}
%%%%%%%%%%%%%%%%%%%%%%%%%%%%%%%%%%%%%%%%%%%%%%%%%%%%%%%%%%%%%%%%%%%%

%\input{iasted03_surProp}
%******************************* SECT 2 *********************************
\secSpc
\section{REPRESENTATION OF QUADRAT-
IC $\;$  HYPER-SURFACE}
%\section{REPRESENTATION OF QUADRA -TIC HYPER-SURFACE}
%\section{Representation of Quadratic Hyper-Surfaces}
$\cal{A}$t first we treat the class of algebraic surfaces in
${\Real}^3$ described by quadratic polynomials. Mercifully, the
corresponding system of transformations (\ref{PQSgen}) can be
linearized. The next step is to determine the boundary of the
regions representing the surface. Without getting into details,
the existence of the boundary can be assured by selecting an
appropriate spacing of axes in the system of the $\|$-coords,
which eventually reflect by changing the constant multipliers of
the first equation in (\ref{PQSgen}).

%~~~~~~~~~~~~~~~~~~~~~~~~~~~~ SUB ~~ SECT 1 ~~~~~~~~~~~~~~~~~~~~~~~~~~~~~
\subSecSpc
\subsection{Definition of the Regions' Boundary }
Let $\sigma$ be a quadratic surface whose representation is the
region $\bar{\sigma}$. The boundary points are those whose every
neighborhood contains both interior and exterior points. For this
case both the transformation (\ref{PQSgen}) and the surface, are
defined by polynomials and hence are differentiable. The basic
properties including continuity are therefore preserved under the
transformation.

Geometrically we rely on the differentiability in finding those
points $\bar{p} \in \bar{\sigma}$ so that we can ``move'' from
$\bar{p}$ in any direction and still remain in region; these are
interior points of $\bar{\sigma}$. Clearly the boundary points are
easily found as the complement of the interior of $\bar{\sigma}$.
The condition for determining whether a point is interior or not
is given by \emph{theorem of implicit function} \cite{CALC85}.
Equivalently, a point $ \bar{a} \in \bar{\sigma}$ is interior
point if and only if the Jacobian, $J(F)|_{ \bar{a}}$, at this
point is different from zero. Conversely, a point $ \bar{b} \in
\bar{\sigma}$ for which this is not true is necessarily a boundary
point; namely, a point $ \bar{b} \in \bar{\sigma}$ s.t. $J(F)|_{
\bar{b}}=0$. In essence the theorem tells us that $\bar{\sigma}$
is closed set and the complement of its interior is the sought
after boundary.

 Generalizing the above for $\Real^{n}$ we get the mapping
 $(x_{1}, \dots ,x_{n}) \rightarrow (\eta,\xi,\psi)$ into the
 projective space, were $x=\frac{\eta}{\psi}$ and
 $y=\frac{\xi}{\psi}$.
 Restating the condition in terms of differential products using
 homogeneous coordinates with the variables $\eta$, $\xi$ and
 $\psi$ yields,
% **** PLEASE EXPLAIN ****
 \begin{equation}\label{J(F)homo}
  (\eta d \xi d \psi - \xi d \eta d \psi + \psi d \eta d \xi)dF=0.
 \end{equation}
 This form is more convenient for handling hyper-surfaces embedded
 in $\Real^{n}$ where $n>3$, for $n=3$ eq. (\ref{J(F)homo}) can be
 written equivalently in term of Jacobian as,
\begin{equation}\label{J(F)}
J(F) \psi ^{3} = Det \left[ \matrix{
  \frac{\partial{F}}{\partial{x_{1}}} & \frac{\partial{F}}{\partial{x_{2}}} & \frac{\partial{F}}{\partial{x_{3}}} \cr
  \frac{\partial{(\eta / \psi)}}{\partial{x_{1}}} & \frac{\partial{(\eta / \psi)}}{\partial{x_{2}}} & \frac{\partial{(\eta / \psi)}}{\partial{x_{3}}} \cr
  \frac{\partial{(\xi / \psi)}}{\partial{x_{1}}} & \frac{\partial{(\xi / \psi)}}{\partial{x_{2}}} &
  \frac{\partial{(\xi / \psi)}}{\partial{x_{3}}} \cr
} \right] \psi ^{3}.
\end{equation}
Substituting $\eta$, $\xi$ and $\psi$ in terms of the $x_{i}$'s
yields an equation which defines an algebraic surface, $\sigma'$,
in $\Real^{3}$. Geometrically, the boundary $\bar{\gamma}$ consist
of points which represent tangent planes touching at points of
$\sigma$ on the intersection $\sigma' \cap \sigma$. Hence,
$\bar{\gamma}$ is the image of the algebraic curve $\sigma' \cap
\sigma = \gamma$ (see Fig. \ref{figGenralCurve}).

%It can be proved that this representation of a quadratic surfaces
%$\sigma$ with some corresponding extension is unique.

%%%%%%%%%%%%%%%%%%%%%%%%%%%%%%%%%%%%%%%%%%%%%%%%%%%%%%%%%%%%%%%%%%%%
%%                     saddle_a
\topFigSpc
\begin{figure*}%[!h]
\centering
% sigma      : z=-(x/a)^2+(y/b)^2
% gamma      : z = -((x/a)^2-x-(y/b)^2+2y)
% bar{gamma} : -(a/2)x^2+(1/2)x-y-(b/2)y^{2}
% parameter  : a=2, b=2
% bar{gamma} : 16-16x-4y+y^{2}-4xy+4x^{2} = 0
\includegraphics[width=\FigWidth in]{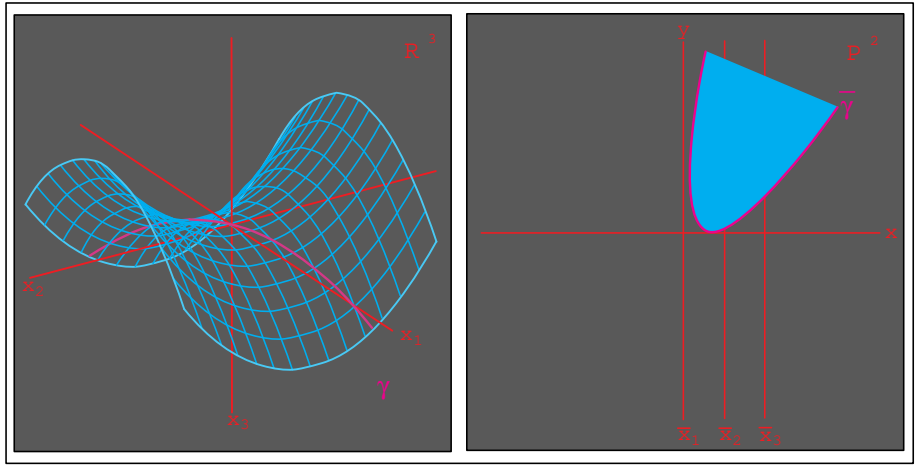}
\capFigSpc \caption{ \label{figSaddle} Saddle $\sigma$:
$z=-(x/2)^2+(y/2)^2 \; \;  \rightarrow \; \;
\bar{\gamma}:16-16x-4y+y^{2}-4xy+4x^{2} = 0$.} \butFigSpc
\end{figure*}
%%%%%%%%%%%%%%%%%%%%%%%%%%%%%%%%%%%%%%%%%%%%%%%%%%%%%%%%%%%%%%%%%%%%

Combining the criterion, eq. (\ref{J(F)}), for the boundary with
the equation of the surface (embedded in 3-dimensional space) and
the transformation equations (in homogenous coordinates) yields :
    \begin{equation}\label{EQsys}
          \begin{array}{lll}
               F(x_{1}, x_{2} ,x_{3}) & = 0, \\
               \eta Q(x_{1}, x_{2} ,x_{3}) - \psi P(x_{1}, x_{2} ,x_{3}) & = 0, \\
               \xi Q(x_{1}, x_{2} ,x_{3}) - \psi S(x_{1}, x_{2} ,x_{3}) & = 0, \\
               J(F(x_{1}, x_{2} ,x_{3})) \psi ^{3} & =
               0.
               \end{array}
          \end{equation}
Solving for $\eta$, $\xi$ and $\psi$ yields the equation of the
boundary. Note that if $F$ is a polynomial of degree $2$, then the
degree of $J(F) \psi ^{3}$ is $\leq 2$ in terms of all variables,
while it is linear in terms of $x_{1}$, $x_{2}$ and $x_{3}$.

Thus far we have constructed a system of four equations
(\ref{EQsys}) in six variables which define a mapping from the
$\Real^3$ into the projective plane $\Proj^2$. Our aim, is to
determine the specific equation of the region's boundary
explicitly. This involves solving this system of equations in
terms of $\eta$, $\xi$ and $\psi$ by eliminating the variables
$x_{1}$, $x_{2}$ and $x_{3}$.

The equation's structure turns out to be very advantageous. Since
the last three equations are linear, the elimination can be done
by isolating a variable (finding an explicit expression in term of
the other variables), and substituting in the remaining linear
equations. When all is said and done, each of the variables
$x_{1}$, $x_{2}$ and $x_{3}$ can be expressed as a rational
equation in $\eta$, $\xi$ and $\psi$. Upon substitution of these
expressions into $F$ the boundary's equation in homogeneous
coordinates is obtained. It follows that the boundary
$\bar{\gamma}$ is a quadratic curve.

%~~~~~~~~~~~~~~~~~~~~~~~~~~~~ SUB ~~ SECT 2 ~~~~~~~~~~~~~~~~~~~~~~~~~~~~~
\subSecSpc
\subsection{Algorithm}
The algorithm's input is an equation of algebraic surface $\sigma
: F(x_{1},x_{2},x_{2}) = 0$ of degree two and the output is the
polynomial which describes the boundary of the surface's image in
$\|$-coords. It is noteworthy that the algorithm applies to
implicit or explicit polynomials with or without singular points.
The formal description is followed by examples which clarify the
various stages and their nuances.

For a given polynomial equation $F(x_{1},x_{2},x_{3})=0$ of degree
2 and a spacing of axes $S_{1 \dots n}$:
\begin{itemize}
  \item Let :

$ \qquad \eta =
\sum_{i=1}^{3}(i-1)\frac{\partial{F}}{\partial{x_{i}}} \;, $

$\qquad  \xi =
\sum_{i=1}^{3}x_{i}\frac{\partial{F}}{\partial{x_{i}}} -2F \; ,$

$\qquad \psi = \sum_{i=1}^{3}\frac{\partial{F}}{\partial{x_{i}}}.
$

\item Write the three linear equations:

$a)$ $\qquad \psi
\sum_{i=1}^{3}(i-1)\frac{\partial{F}}{\partial{x_{i}}} -
   \eta \sum_{i=1}^{3}\frac{\partial{F}}{\partial{x_{i}}} = 0,$

$b)$ $ \qquad \psi
(\sum_{i=1}^{3}x_{i}\frac{\partial{F}}{\partial{x_{i}}} -2F) -
   \xi \sum_{i=1}^{3}\frac{\partial{F}}{\partial{x_{i}}} =0, $

$c)$ $ \qquad J(F) \psi ^3 = 0.$

  \item Using substitution write

$ \qquad x_{i} =f_{i}(\eta,\xi,\psi) \;$, $\; \; \;$  for
$i=1,2,3$.

  \item Substitute

  $ \qquad F(f_{1}(\eta,\xi,\psi),f_{2}(\eta,\xi,\psi),f_{3}(\eta,\xi,\psi))=0.$

  \item Retain the equation's numerator.

  \item The output is obtained by substitution:

  $ \qquad \eta \leftarrow x \; \; , \; \; \xi \leftarrow y \; \; , \; \; \psi \leftarrow 1.$
\end{itemize}
All this falls into place with the following examples.

%%%%%%%%%%%%%%%%%%%%%%%%%%%%%%%%%%%%%%%%%%%%%%%%%%%%%%%%%%%%%%%%%%%%
%%                     sphere_a
\topFigSpc
\begin{figure*}%[!h]
\centering
% sigma      : (a*x)^2 + (b*y)^2 + (c*z)^2 = d
% gamma      : z = -((x/a)^2-x-(y/b)^2+2y)
% bar{gamma} :
% parameter  : a=2, b=2 c=1 d=2
% bar{gamma} : x^2-4xy+y^2+1=0
\includegraphics[width=\FigWidth in]{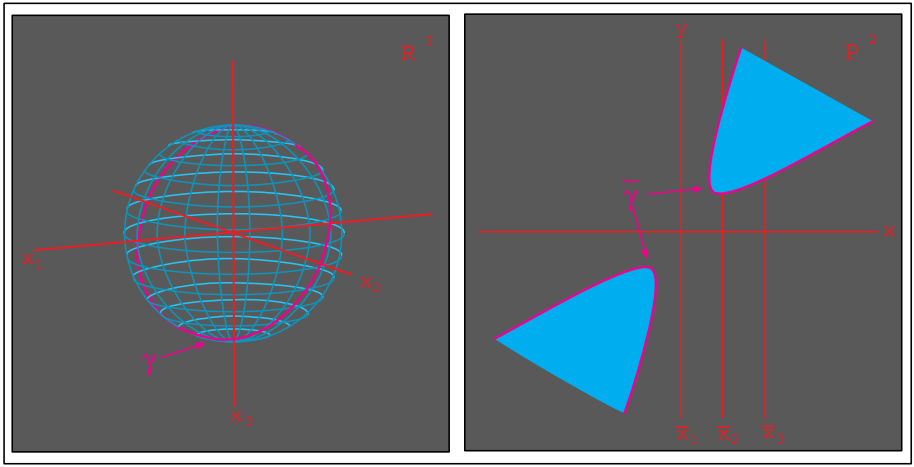}
\capFigSpc \caption{\label{figShepre} Sphere $\sigma$: $x^2 + y^2
+ z^2 = 2 \; \; \rightarrow \; \; \bar{\gamma}: x^2-4xy+y^2+1=0$.}
\butFigSpc
\end{figure*}
%%%%%%%%%%%%%%%%%%%%%%%%%%%%%%%%%%%%%%%%%%%%%%%%%%%%%%%%%%%%%%%%%%%%

%%%%%%%%%%%%%%%%%%%%%%%%%%%%%%%%%%%%%%%%%%%%%%%%%%%%%%%%%%%%%%%%%%%%
%%                     hyper1_a
\topFigSpc
\begin{figure*}%[!h]
\centering
% sigma      : (a*x)^2 + (b*y)^2 - (c*z)^2 = d
% gamma      : x-2y+z=0
% bar{gamma} : u^{2}a_{1}+2uva_{4}+2uwa_{5}+v^{2}a_{2}+2vwa_{6}+a_{3}w^{2}
% parameter  : a=1, b=1 c=1 d=1
% bar{gamma} : x^2-4xy+y^2-1=0
\includegraphics[width=\FigWidth in]{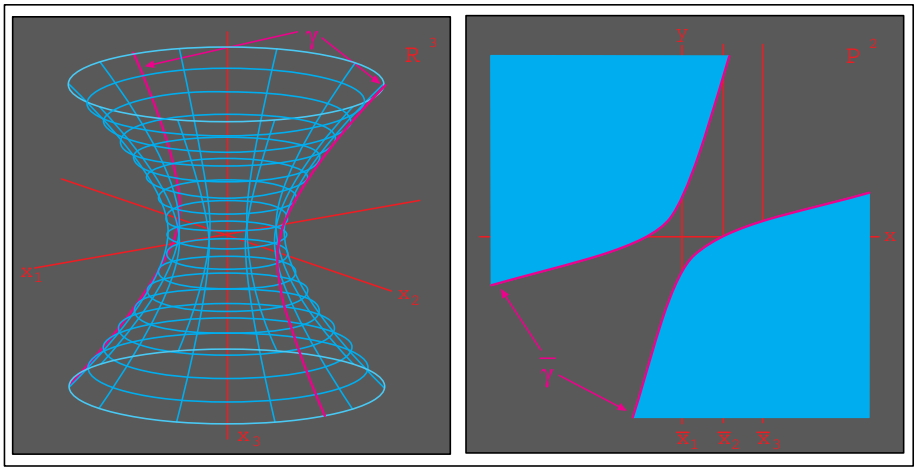}
\capFigSpc \caption{  \label{figHyperOne} Hyperboloid of one sheet
$\sigma$ : $x^2 + y^2 - z^2 = 1 \; \;  \rightarrow \; \;
\bar{\gamma}: x^2-4xy+y^2-1=0$.} \butFigSpc
\end{figure*}
%%%%%%%%%%%%%%%%%%%%%%%%%%%%%%%%%%%%%%%%%%%%%%%%%%%%%%%%%%%%%%%%%%%%

%%%%%%%%%%%%%%%%%%%%%%%%%%%%%%%%%%%%%%%%%%%%%%%%%%%%%%%%%%%%%%%%%%%%
%%                     hyper2_a
\topFigSpc
\begin{figure*}%[!h]
\centering
% sigma      : (a*x)^2 + (b*y)^2 - (c*z)^2 = d
% gamma      : x-2y+z=0
% bar{gamma} : $u^{2}a_{1}+2uva_{4}+2uwa_{5}+v^{2}a_{2}+2vwa_{6}+a_{3}w^{2}$
% parameter  :
% bar{gamma} :x^2-2xy+4y^2-1=0
\includegraphics[width=\FigWidth in]{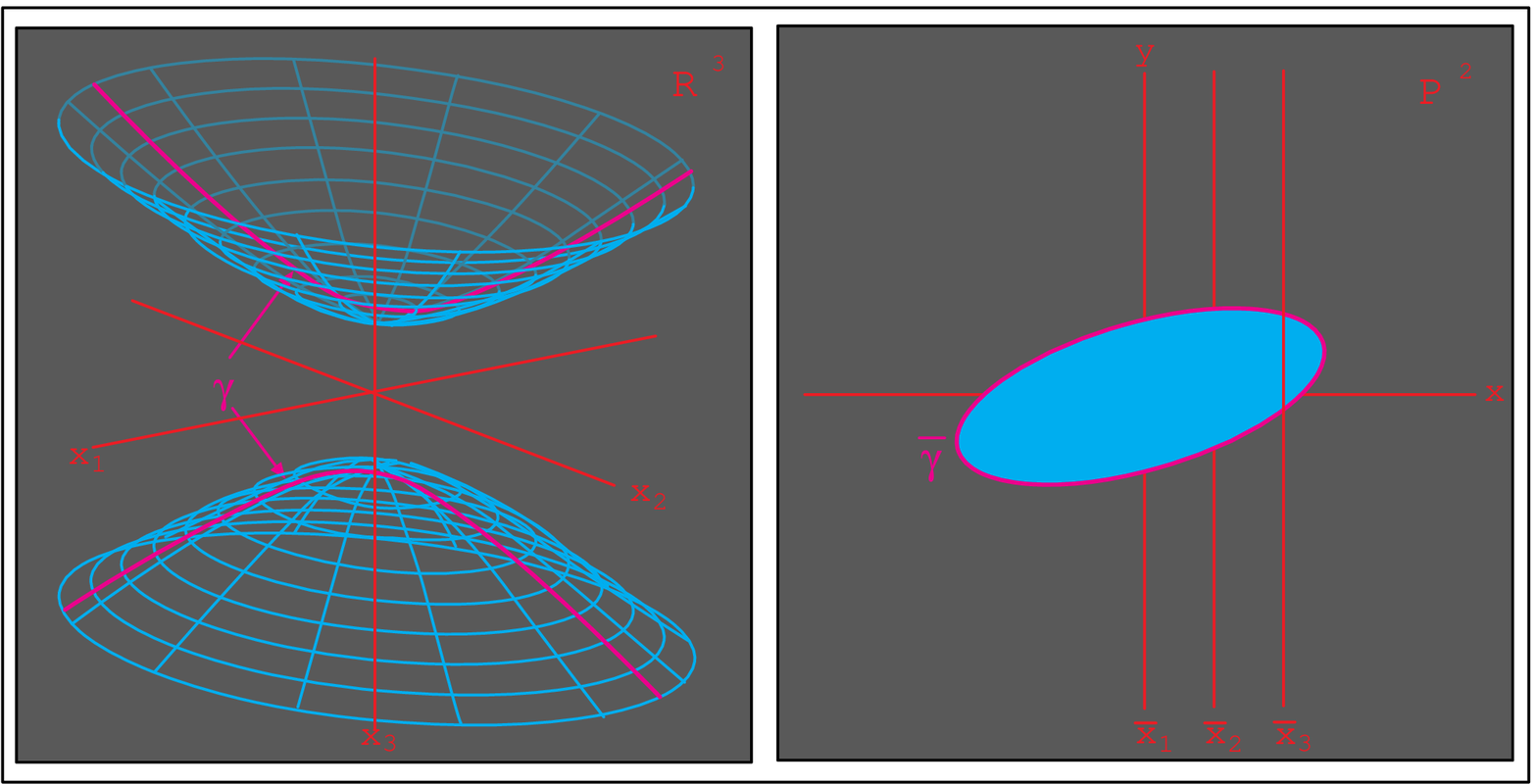}
\capFigSpc \caption{\label{figHyperTwo} Hyperboloid of two sheets
$\sigma$: $x^2-4y^2+2z^2=-2 \; \;  \rightarrow \; \; \bar{\gamma}:
x^2-2xy+4y^2-1=0$.} \butFigSpc
\end{figure*}
%%%%%%%%%%%%%%%%%%%%%%%%%%%%%%%%%%%%%%%%%%%%%%%%%%%%%%%%%%%%%%%%%%%%

%******************************* SECT 3 *********************************
\secSpc
\section{EXAMPLE OF QUADRATIC SURFACE AND THEIR TRANSFORMS}
%\section{Examples of Quadratic Surfaces and their Transforms}

$\cal{I}$n the first example is quite detailed to accommodate the
readers wishing to follow the application of the algorithm in
depth. Let $\sigma$ be $3$-dimensional saddle (see Fig
\ref{figSaddle}) generated by the polynomial equation,

$F(\bar{x})=\left( \frac{x_{1}}{2}\right) ^{2}-\left(
\frac{x_{2}}{2}\right) ^{2}+x_{3}=x_{1}^{2}-x_{2}^{2}+4x_{3}=0$,
where $\bar{x} = (x_{1},x_{2},x_{3})$, and $S_{1 \dots n}$ the
standard spacing of axes.

\begin{description}
  \item[step 1] Let :

$\begin{array}{llll} P(\bar{x}) = &
\sum_{i=1}^{3}(i-1)\frac{\partial{F}}{\partial{x_{i}}} & = &
-2x_{2}+8, \\
S(\bar{x}) = &
\sum_{i=1}^{3}x_{i}\frac{\partial{F}}{\partial{x_{i}}} -2F & = &
-4x_{3}, \\
Q(\bar{x}) = & \sum_{i=1}^{3}\frac{\partial{F}}{\partial{x_{i}}} &
= & 2x_{1}-2x_{2}+4.
\end{array}$

%$\eta= P = \sum_{i=1}^{3}(i-1)\frac{\partial{F}}{\partial{x_{i}}}
%\; \; =-2x_{2}+8$ $\qquad \; \; \; \; \; \;  \xi = S =
%\sum_{i=1}^{3}x_{i}\frac{\partial{F}}{\partial{x_{i}}}
%-2F=-4x_{3}$ $\qquad \; \; \; \; \;   \psi = Q =
%\sum_{i=1}^{3}\frac{\partial{F}}{\partial{x_{i}}} \qquad \; \; \;
%\; \; =2x_{1}-2x_{2}+4$
  \item[step 2]
Write\footnote{The authors acknowledge and are grateful for the
use of the symbolic manipulation program {\em Singular} developed
by the Algebraic Geometry Group, Department of Mathematics,
University of Kaiserslautern, Germnay.} the three linear
equations:

$a)$ $\eta Q(\bar{x}) - \psi P(\bar{x})= $

$\qquad \qquad 2( \eta x_{1} +(\psi - \eta) x_{2}+2 \eta - 4 \psi)
=0$,

$b)$ $\xi Q(\bar{x})- \psi S(\bar{x})= $

$\qquad \qquad 2( \xi x_{1}- \xi x_{2}+2 \psi x_{3}+2 \xi) =0$,

$c)$ $ J(F(\bar{x})) \psi ^ 3 = (x_{1} (\psi - \eta) +x_{2} \eta -
\xi)$.

Notice: substitution of $\eta$, $\xi$ and $\psi$ in terms of
$x_{1}$, $x_{2}$ and $x_{3}$ yields a surface in $\Real^{3}$,

 $  \qquad \sigma' :
32(-x_{2}^{2}+4x_{2}+x_{1}^{2}+2x_{3}-2x_{1}) =0$.

Hence $\gamma = \sigma' \cap \sigma $,

$\qquad \gamma : \; \; \left\{ \begin{array}{llll}
               x_{1}^{2}-x_{2}^{2}+4x_{3}  & = 0 \\
               -x_{2}^{2}+4x_{2}+x_{1}^{2}+2x_{3}-2x_{1} & = 0
               \end{array}
           \right.$.

  \item[step 3]  Using simple substitution write

  $b)$ $ \Rightarrow \;x_{3}=-\frac{\xi ( x_{1}-x_{2}+2) }{2\psi }$,

  $c)$ $ \Rightarrow x_{2}=-\frac{(\psi - \eta) x_{1} - \xi
  }{\eta}$.

Then using equation $a)$ we get,

 $ \qquad x_{1}= -\frac{-2 \eta^{2}+ \eta \xi - \psi \xi +4\psi
\eta}{\psi \left( \psi -2 \eta \right) },$

$ \qquad x_{2}= \frac{2\eta ^{2}- \eta \xi -6 \psi \eta +4 \psi
^{2}}{\psi \left( \psi -2 \eta \right) },$

$\qquad x_{3}= \frac{\xi(- \xi +2 \eta + 2 \psi )}{2 \psi( \psi -2
\eta)}$.

  \item[step 4] Substitute in $F$,

$x_{1}^{2}-x_{2}^{2}+4x_{3}=$

$\qquad -\frac{16 \psi ^{2}-16 \psi \eta -4 \psi \xi + \xi ^{2}-4
\eta \xi +4 \eta ^{2}}{\psi \left( \psi -2 \eta \right) }=0$.

  \item[step 5] Retain the equation's numerator,

  $16 \psi ^{2}-16 \psi \eta -4 \psi \xi + \xi ^{2}-4 \eta \xi +4
\eta ^{2} = 0$.

  \item[step 6] Finally, the output is obtained by substitution,

  $  \eta \leftarrow x \; \; , \; \; \xi \leftarrow y \; \; , \; \; \psi \leftarrow
  1$:

$\bar{\gamma} : \; \;  16-16x-4y+y^{2}-4xy+4x^{2} = 0.$

\end{description}

\ni The surface and its image including the boundary curve are
shown in Fig \ref{figSaddle}. The representation of other
quadratic surfaces is illustrated in figures \ref{figShepre},
\ref{figHyperOne} and \ref{figHyperTwo}.

%******************************* SECT 4 *********************************
%\secSpc
%\section{GENERALIZATION FOR MULTIDIMENSIONAL QUADRATICS HYPER-SURFACES}
%\section{Generalization for Multidimensional Quadratics Hyper-Surfaces}

%$\cal{S}$ummarizing, the representation of 3-dimensional quadratic
%surfaces is obtained from :
%\begin{enumerate}
% \item{the criterion for the boundary points, and}
% \item{solving the system of equations.}
%\end{enumerate}
%For dimension $n = 3$ there are only two indexed regions. The
%generalization to arbitrary dimension $n$, to be covered in a
%future paper, develops along similar lines.

%The boundary criterion, (\ref{J(F)homo}), in homogenous
%coordinates is defined for any dimension. The following part
%involves the elimination requiring that the number of equations be
%greater then the number of variables. Hence the system needs to be
%extended to at least $n+1$ equations. Without entering into the
%details it is possible to obtain the additional equations to
%complete the construction algorithm. Then the hyper-surface's
%representation consists of $n-1$ regions stemming from the $n-1$
%indexed points representing each of the tangent hyper-planes.

%******************************* SECT 4 *********************************
\secSpc
\section{CONCLUSION}
%\section{Conclusions}

$\cal{T}$he new representation

\begin{itemize}
 \item{is constructive,}
 \item{enables the representation of non-convex objects,}
 \item{maps algebraic surfaces to regions having algebraic
  curves as boundaries,}
% \item{provides a new basis for the approximation of complex surfaces, from their
% corresponding {\em planar} regions.}
\end{itemize}
An important ``fringe benefit'' is the avoidance of the
``over-plotting'' problem in $\|$-coords where polygonal lines
obscure portions of the display.

%\pagebreak


\begin{thebibliography}{10}

\bibitem[Cox97]{cox97ideals}
{\sc Cox, D., Little, J., and O'Shea, D.}
\newblock 1997.
\newblock {\em Ideals, Varieties, and Algorithms}, second ed.~ed.
\newblock Springer, New York.

\bibitem[Dim84]{dimsdale84conic}
{\sc Dimsdale, B.}, 1984.
\newblock Conic transformations and projectivities.
\newblock IBM Los Angeles Scientific Center.
\newblock Rep. G320-2753.

\bibitem[Har92]{harris92algebraic}
{\sc Harris, J.}
\newblock 1992.
\newblock {\em Algebraic Geometry}, vol.~A rst course.
\newblock Springer, NY.

\bibitem[Hun92]{hung92smooth}
{\sc Hung, C., and Inselberg, A.}, 1992.
\newblock Parallel coordinates representation of smooth hypersurfaces.
\newblock IBM Palo Alto Scientific Center.
\newblock Tech Rep. G320-3575.

\bibitem[Ins85]{inselberg85plane}
{\sc Inselberg, A.}
\newblock 1985.
\newblock The plane with parallel coordinates.
\newblock {\em The Visual Computer 1}, 2, 69--92.
%\pagebreak

\bibitem[Ins99]{Insel99a}
{\sc Inselberg, A.}
\newblock 1999.
\newblock Don't panic ... do it in parallel!
\newblock {\em Computational Statistics 14\/}, 53--77.

\bibitem[Izh01]{zur:01:ms}
{\sc Izhakian, Z.}
\newblock 2001.
\newblock {\em An Algorithm For Computing A Polynomial's Dual Curve In Parallel
  Coordinates}.
\newblock M.sc thesis, University of Tel Aviv.

\bibitem[Mar85]{CALC85}
{\sc Marsden, J., and Weinstein, A.}
\newblock 1985.
\newblock {\em Calculus}.
\newblock Springer-Verlag, New York.

\end{thebibliography}
\end{document}